\documentclass[prl,twocolumn,superscriptaddress,showpacs,floatfix]{revtex4}
\usepackage{epsfig}
\usepackage{graphicx}
\usepackage{dcolumn}
\usepackage{bm}

\begin{document}

\title{An all-optical event horizon in an optical analogue of a Laval nozzle}

\author{M. Elazar}
\author{V. Fleurov}
\author{S. Bar-Ad}

\affiliation{Raymond and Beverly Sackler School of Physics and Astronomy,
Tel-Aviv University, Tel-Aviv 69978, Israel}

\date{\today}

\begin{abstract}

Exploiting the fact that light propagation in defocusing nonlinear
media can mimic the transonic flow of an equivalent fluid, we
demonstrate experimentally the formation of an all-optical event
horizon in a waveguide structure akin to a hydrodynamic Laval
nozzle. The analogue event horizon, which forms at the nozzle throat
is suggested as a novel platform for analogous gravity experiments.

\end{abstract}


\maketitle


Event horizons are well known in the context of astrophysics and
cosmology. Less known are the analogy between an astrophysical event
horizon and the sonic horizon in transonic fluid flow, and the
prediction that a thermal spectrum of sound waves should be emitted
from a sonic horizon, in analogy with Hawking radiation
\cite{Unruh}. These analogies set the stage for attempts to create
laboratory black hole analogues, involving various physical
scenarios, from water flowing in a channel to the acceleration of a
superfluid to nonlinear optical experiments
\cite{Rousseaux1,Philbin,Marino,Recati,Lahav,Rousseaux2,Belgiorno,Fouxon,Weinfurtner}.
We have recently proposed an alternative approach to analogue
gravity experiments -- an all-optical experiment based on laser
light propagation in a distinctive nonlinear waveguide, which is
analogous to a Laval nozzle (a well-known device in the context of
aerodynamics). This approach has two great advantages over previous
experiments: The attainment of supersonic velocities is very easy,
and the analogue of Hawking radiation has a unique optical
signature, which can be readily detected \cite{Fouxon}. The analogy
is based on the realization that under certain conditions light can
"flow" through certain types of media in a fashion reminiscent of
actual fluid flow. A prime example is a laser beam propagating
through a Kerr-type nonlinear medium, which is usually described
analytically by the Non-Linear Schr\"{o}dinger  equation
\cite{Zakharov}. The latter can be mapped, through the Madelung
transformation \cite{Madelung}, to a pair of coupled equations for
the amplitude and phase, which have the form of continuity and Euler
equations for an equivalent fluid
\cite{Marino,Fouxon,Madelung,Marburger,El,Fleischer,Jia,Barsi},
which may be called "luminous fluid":

\begin{eqnarray}\label{hydrodynamics}
&& \partial_z\rho + \nabla\cdot[\rho{\bm v}] = 0,\\
&& \partial_z\bm v + \frac{1}{2} \nabla{\bm v}^2 = -
\frac{1}{\beta_0}\nabla\left(- \frac{1}{2\beta_0} \frac{\nabla^2
f}{f} + U_{\it ext} + \lambda \rho \right).\nonumber \label{b1}
\end{eqnarray}

Eqs. (\ref{hydrodynamics}) describe the evolution of the complex
amplitude $f(x,y) e^{-i\varphi(x,y)}$ as the light propagates along
the $\emph{z}$ axis with the wave vector $\beta_0$. Here $\rho =
f^2$ is the light intensity, and the transverse component of the
wave-vector, $\beta_0 \bm v = -\nabla\varphi$, plays the role of
velocity. The coordinate $\emph{z}$ plays the role of time,
$\beta_0$ is equivalent to the mass of a particle, and the spatially
inhomogeneous refraction index assumes the role of a potential,
$U_{ext}(x,y)$. The nonlinear term is due to the Kerr effect. Thus
incident light, which propagates at an angle relative to the
$\emph{z}$ axis is mapped onto a fluid with a finite transverse
velocity, and a change of that angle corresponds to acceleration of
the fluid. This approach has proved to be an extremely powerful one
when applied to the problem of coherent tunneling
\cite{Fleurov,Dekel1,Dekel2,Barak,Wan,Cohen}. It has also been used
to model dispersive shock waves that appear when the nonlinearity is
repulsive (\emph{i.e.} self-defocusing, $\lambda>0$), and
consequently an equivalent real sound velocity
$s^2=\lambda\rho/\beta_0$ can be defined
\cite{El,Fleischer,Jia,Barsi,Hakim,Leszczyszyn,Gladush}. Our
proposal for an optical analogue of the Laval nozzle uses a similar
approach,\cite{Fouxon,Fleurov2} and shows that when the light is
confined by a properly shaped channel (\emph{i.e.} waveguide),
represented by the transverse potential $U_{ext}$, it may propagate
transversally in a way that resembles the accelerating transonic
flow of the equivalent, luminous fluid. Thus low-incidence,
(\emph{i.e.} "subsonic") laser light is predicted to accelerate,
(\emph{i.e.} change its propagation direction) while traversing the
nozzle, reaching a critical velocity, which is equivalent to the
sound velocity in a real fluid, at the nozzle throat, and exiting
the nozzle at a "supersonic" velocity. While the notion that light
propagation in nonlinear media can mimic the flow of an equivalent
fluid is not new, the equivalent of transonic acceleration in an
optical analogue of a Laval nozzle has never been observed before in
an actual experiment, and is demonstrated here for the first time.

\begin{figure}
\begin{center}
\includegraphics[width=8.0cm,angle=0]{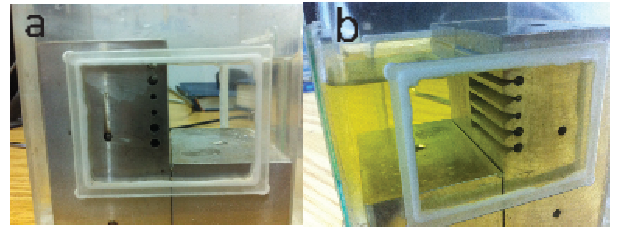}
\caption{(Color online) Images of the waveguide structures. (a) shows the input
plane, with six circular openings of lightpipes of different
dimensions. (b) shows the exit plane, with grooves of triangular
cross-section, cut half-way along the sides of the channels, and
forming the divergent sections of the nozzles.} \label{Cell}
\end{center}
\end{figure}

\begin{figure}
\begin{center}
\includegraphics[width=8.0cm,angle=0]{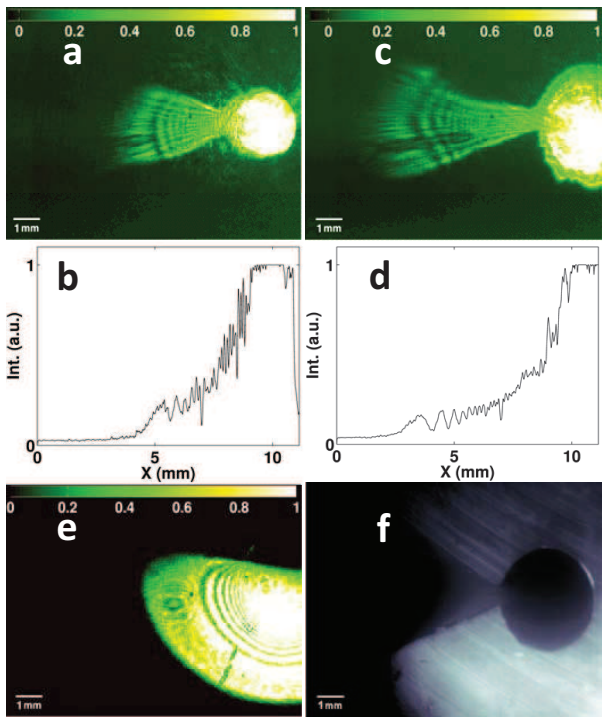}
\caption{(Color online) Waveguide exit plane images and corresponding power density
cross-sections along the nozzle axis for an input power of 2 Watts
and an iodine concentration of $ \sim$40 ppm. (a) and (b) show data
for a 2 mm diameter waveguide, (c) and (d) show data for a 3 mm
diameter waveguide, and (e) shows the free expansion of the beam
outside the waveguide structure. (f) is a reference image of the 3
mm diameter waveguide.} \label{Figure1}
\end{center}
\end{figure}

We study the flow of luminous liquid through an optical Laval nozzle
experimentally by launching a continuous wave laser beam into an
appropriately shaped waveguide with reflective walls, filled with a
Kerr-type defocusing nonlinear material. The experimental challenge
here is to create conditions of steady flow with a subsonic input
velocity. Since the velocity is $k_x/\beta_0$ (where $k_x$ is the
$x$-component of the wave vector) and the sound velocity is $s^2=
\lambda \rho/\beta_0$, the above input conditions imply a small
input angle of the beam and a high nonlinearity and/or input
intensity (Note that, for a given angle, low and high intensities
correspond to supersonic and subsonic flow, respectively). However,
an unavoidable consequence of these conditions is strong
self-defocusing of the beam, and as a result the wave packet, which
traverses the nozzle is an expanding "droplet" of liquid, with a
tendency of the power density in the cavity to decrease with
increasing input power. Furthermore, while the peak intensity of the
droplet may correspond to subsonic flow, it is always surrounded by
supersonic flow (in contrast to the usual case in hydrodynamics),
and when confined to a Laval nozzle such as the one discussed in
Ref. \onlinecite{Fouxon}, the fluid flows from the throat towards
both sides of the nozzle. It is thus impossible to generate the
steady sonic background flow conditions stipulated by the theory in
a simple waveguide with a convergent-divergent cross-section formed
by two convex walls. To circumvent this problem we use an
alternative waveguide design, based on a light-pipe of circular
cross-section drilled in an aluminum block, with a groove of
triangular cross-section, cut along the side of the channel, acting
as the divergent section of the nozzle (see Fig. \ref{Cell}). The
total length of each light pipe is $L$ = 67 mm, and the groove
extends over the second half of this length. This design is intended
to "trap" the expanding beam and confine it in a homogeneous,
high-density and low-velocity mode, thus preparing it for ejection
through the groove, and is akin to the configuration of a rocket
engine: a high-pressure gas is first loaded into a combustion
chamber, and is then expelled through the nozzle. The aluminum
block, with several nozzles of different diameters, aperture sizes
and groove opening angles, is enclosed in a plexiglass cell with
glass windows, which is sealed and filled with iodine-doped ethanol.
The nonlinear index variations result from optical absorption by the
iodine, which in turn leads to thermally-induced changes of the
index of refraction -- a non-local nonlinearity, which slightly
washes out the thermal gradients \cite{Barsi}. The nonlinearity
$\lambda\rho$  can be expressed, in terms of the nonlinearly-induced
refractive index change $\delta n$, as $\delta n\beta_0/n_0$, where
$n_0$ is the linear refractive index of the material \cite{Barsi}.
The corresponding dimensionless sound velocity is then $s^2 = \delta
n/n_0$, meaning that the input beam is subsonic for $k_x/ \beta_0 <
\sqrt{\delta n/n_0}$. We use a continuous-wave frequency-doubled YAG
laser (532 nm), and focus the beam to a $\sim$ 0.5 mm waist at the
input of a waveguide. The input power is varied by means of the
laser controller, in order to avoid thermal effects in
variable-density filters. Images of the exit plane of the waveguide
are recorded by means of a CCD camera. In all cases images were
acquired after stabilization of the thermal gradients.

\begin{figure}
\begin{center}
\includegraphics[width=8.4cm]{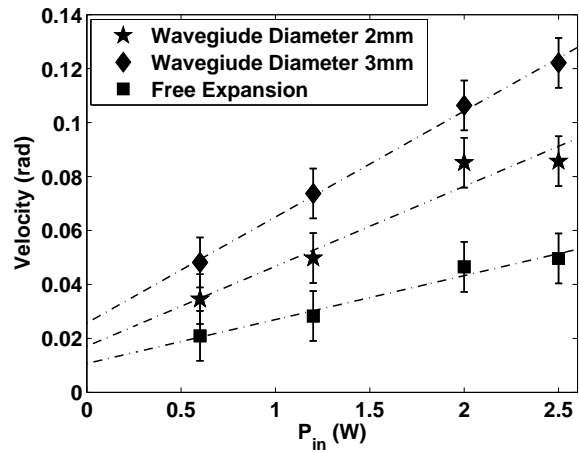}
\caption{The measured jet velocities as a function of input power
for an iodine concentration of $ \sim$ 40 ppm. Data are shown for
two waveguides and for free expansion of the beam outside the
waveguide structure. The calculation of velocities is explained in
the text. The lines are guides to the eye.} \label{Figure2}
\end{center}
\end{figure}

Figure \ref{Figure1} presents images of the exit plane of two of the
waveguides, and the corresponding power density cross-sections along
the nozzle axis, for an input power (2 Watts) that is sufficiently
high to completely fill the waveguides (at an iodine concentration
of $ \sim$ 40 ppm). Figure \ref{Figure1}(a) shows a 2 mm diameter
waveguide, and Fig. \ref{Figure1}(c) shows a 3 mm diameter
waveguide, both having a $\sim$ 0.5 mm opening (\emph{i.e.} nozzle
throat). Figures \ref{Figure1}(b) and \ref{Figure1}(d) are the power
density cross-sections corresponding to Figs. \ref{Figure1}(a) and
\ref{Figure1}(c), respectively, obtained by summation over 12 CCD
lines at the center of each nozzle. Figure \ref{Figure1}(e) shows
the free expansion of the beam when it propagates outside the
waveguide structure, and Fig. \ref{Figure1}(f) shows a reference
image of the 3 mm diameter waveguide, obtained with incoherent light
and with the laser beam blocked. Figures \ref{Figure1}(a)-(d)
clearly show the jets of luminous liquid ejected from the nozzles as
the beam propagates through the waveguides. Note that the jets
extend farther than the edge of the beam undergoing free expansion
(Fig. \ref{Figure1}(e) -- A detailed analysis is presented in Fig.
\ref{Figure2}). Furthermore, there is a sharp drop in the density as
the jet exits the nozzles, which is clearly seen in the images and
in the power density cross-sections. This demonstrates that the
luminous liquid is accelerating at the nozzle throat rather than
gradually expanding through the opening. Finally, while the confined
beam propagates along the waveguide walls at a very slow
(\emph{i.e.} subsonic) velocity, the following analysis shows that
the jet of luminous liquid is indeed supersonic: The dimensionless
velocity of the jet outside the waveguide is first calculated from
its extension in the transverse direction, deduced from the images.
The relation is simply $\bm v = \Delta x/\Delta z = 2\Delta x/L
\sim0.1$, where $\Delta x$ is the transverse distance from the
nozzle throat to the edge of the jet, and $\Delta z = L/2$ is the
distance along the $z$ axis that the same part of the jet has
propagated by the time it reached the exit plane. This velocity
should be compared to the local sound velocity, which can be
estimated by analyzing the light intensity distribution in the exit
plane and the rate of expansion of the freely-expanding (\emph{i.e.}
self-defocusing) beam. The latter, deduced from Fig.
\ref{Figure1}(e), allows us to calculate $\lambda \rho_0$ and the
corresponding sound velocity at the input. The former in turn allows
us to deduce the sound velocity, which corresponds to the lower
density of the jet, taking into account the expansion of the beam in
the light-pipe, the relative intensities of the jet and inside the
light-pipe, and measured losses. This calculation gives a local
sound velocity in the jet on the order of $1 \times 10^{-3}$ or
less, meaning that the local Mach number is $>$100. This clearly
establishes that the luminous liquid undergoes transonic
acceleration and forms a "sonic" horizon as it expands through the
nozzle.

In Fig. \ref{Figure2} the dimensionless velocities ($\bm v = 2\Delta x/L$)
of the jets emanating from the 3 mm and 2 mm nozzles are plotted
as a function of the input intensity, for a fixed iodine
concentration. Also shown is the velocity at the
envelope of the freely-expanding beam, which we estimate as $dx/dz
\approx \Delta x'/L$ (In this case we measure $\Delta x'$ from the
center of the beam, which we determine from low-intensity
measurements; This velocity corresponds to the asymptotic expansion
angle, obtained for $L>>1/\lambda \rho_0$, \emph{i.e.} when the
propagation distance is much longer than the characteristic
defocusing distance, and is a reasonable estimate for intensities $
>$ 1 Watt). Figure \ref{Figure2} clearly shows that the velocity of
the jet is higher than that of the freely-expanding beam throughout
the experimental intensity range, in spite of the fact that the
initial conditions for the free expansion involve higher pressures
(the equivalent of pressure in a luminous liquid is $P =
\frac{1}{2}\lambda \rho^2$). On the other hand Fig. \ref{Figure2}
shows that the jet emanating from the 2 mm waveguide is slower than
that ejected from the 3 mm nozzle, although, for a given input
intensity, the pressure in the latter is supposed to be lower. This
discrepancy can result from the fact that the opening in the 2 mm
waveguide subjects a larger angle, resulting in a less directional
jet (compare Figs. \ref{Figure1}(a) and \ref{Figure1}(c)). We also
measure higher losses (due to scattering and absorption) in the
smaller waveguide, so in fact the power densities in the two
waveguides are comparable. Finally, the non-locality of the
nonlinearity may have a stronger effect on the 2 mm waveguide.

\begin{figure}
\begin{center}

  \includegraphics[width=8.0cm,angle=0]{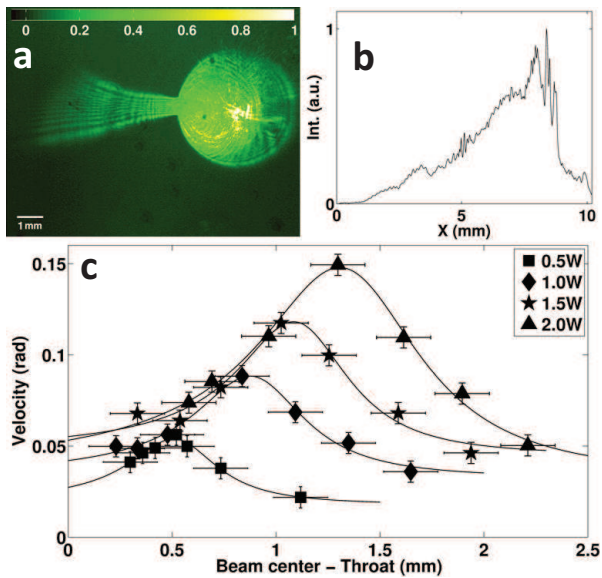}
\caption{(Color online) Acceleration of supersonic flow. (a) An image of the exit
plane of the 3 mm waveguide. (b) The corresponding power density
cross-section along the nozzle axis. (c) The jet velocity as a
function of the distance between the beam axis and the nozzle
throat, for four input powers (The curves are guides to the eye).
The iodine concentration is $ \sim$ 20 ppm.} \label{Figure3}
\end{center}
\end{figure}

Measurements at lower intensities illustrate another regime of
operation of the nozzles. Figure \ref{Figure3}(a) shows an image of
the 3 mm waveguide for self-defocusing (\emph{i.e.} beam expansion)
that is not sufficiently strong to completely fill it (an input
power of 2 Watts and an iodine concentration of only $\sim$20 ppm).
Figure \ref{Figure3}(b) shows the corresponding power density
cross-sections along the nozzle axis. In order to produce the jet
seen in Fig. \ref{Figure3}(a) the nozzle throat had to be displaced
(horizontally) relative to the beam axis. Figure \ref{Figure3}(c)
shows the dependence of the jet velocity on the displacement and the
input power. Note that as the input power increases the optimum
acceleration is obtained when the beam axis is moved farther away
from the throat (at an input power of 2 Watts the beam axis is then
near the center of the waveguide). A comparison of the data in
\ref{Figure3}(c) with the divergences of the freely-expanding beam,
measured separately for the same input powers, shows that optimum
acceleration at the nozzle is obtained when the envelope of the
freely-expanding beam coincides with the nozzle throat halfway
through the waveguide (\emph{i.e.} at $z=L/2$). Under these
conditions the luminous liquid entering the nozzle is already
supersonic, and is accelerated further in the divergent section of
the nozzle. The smooth power density cross-section shown in Fig.
\ref{Figure3}(b) supports this interpretation (compare this to the
sharp density gradients at the throat in Fig. \ref{Figure1}). In
this case the nozzle operates in a regime that is not typical of a
Laval nozzle, though.

In conclusion, we demonstrate experimentally, for the first time,
the transonic acceleration of a luminous liquid through an optical
analogue of a Laval nozzle. The analogue of a sonic event horizon,
which forms at the nozzle throat, lends itself to studies of
classical (and possibly quantum) fluctuations that are akin to
Hawking radiation \cite{Fouxon}. Compared to other experiments and
proposals for "analogue gravity"
\cite{Rousseaux1,Philbin,Marino,Recati,Lahav,Rousseaux2,Belgiorno,Weinfurtner},
our experiment has the advantage that it allows easy generation of
supersonic flow conditions. While the nonlinearity length $l_{nl}=1/
\sqrt{\lambda \rho_0 \beta_0} \sim 30 \mu m$ in our experiment is
sufficiently short to give way to fluctuations with a linear
dispersion relation,\cite{Fouxon} the challenge is to create an
equivalent Hawking temperature that is high enough to measure
experimentally. Note that this is not a real temperature, but it
rather a constraint on the minimum light intensity (and sound
velocity) required in the waveguide:  As explained in Ref.
\onlinecite{Fouxon}, the ratio of amplitudes of the two parts of a
classical fluctuation with wave vector $\nu_0$ -- the part which is
carried away with the supersonic flow (\emph{i.e.} "falls" into the
black hole) and the part which penetrates into the subsonic region
(\emph{i.e.} "escapes" from the black hole), is $\exp(\frac{2\pi c
\nu_0}{\sqrt{3} \overline{s}}) > 1$, where $c$ is a characteristic
length scale of the nozzle and $\overline{s}$ is the sound velocity
at the throat. This ratio needs to be of order unity so that both
parts would be observed in the experiment, and allow direct
measurement of the Hawking temperature $T_H$. The same condition can
also be written as $l_H \approx 2 l_0$, where $l_H = \hbar / T_H = 4
\pi c /\sqrt{3} \overline{s}$ and $l_0 = 1/\nu_0$ are characteristic
length scales of the horizon and the fluctuations, respectively. In
the experiment described here $l_H$ is on the order of a few meters
($c \approx 1\times 10^{-3}$ m, $\overline{s} \approx
1\times10^{-3})$, while $l_0$ must be on the order of a few
centimeters (the length of the cavity). This may still allow
observation of the part of a fluctuation which is carried away with
the supersonic flow, but the part which penetrates into the subsonic
region will most likely be submerged in noise. Therefore $l_H$ must
be decreased by two orders of magnitude. Note, however, that for
given input intensity and nonlinear coefficient the factor
$c/\overline{s}$ in the expression for $l_H$ grows like $c^2$.
We therefore estimate that an order of magnitude smaller cavity will
be sufficient for observing both parts of a straddled classical
fluctuation. The requirement for a slow rate of acceleration can be
met by a refined, smoother waveguide cross-section, compared to the
rudimentary prototype that we have used here for demonstration
purposes.

We thank Mr. G. Elazar for his technical support. This research was
supported by the US-Israel Binational Science Foundation and by the
Israel Science Foundation.


%


\begin{thebibliography}{99}


\bibitem{Unruh} W. G. Unruh, Phys. Rev. Lett. \textbf{46}, 1351 (1981).

\bibitem{Rousseaux1}    G. Rousseaux \emph{et al.}, New J. Phys.
\textbf{10}, 053015 (2008).

\bibitem{Philbin}   T.G. Philbin \emph{et al.}, Science
\textbf{319}, 1367 (2008).

\bibitem{Marino}    F. Marino, Phys. Rev. A \textbf{78}, 063804
(2008).

\bibitem{Recati}    A. Recati, N. Pavloff, and I. Carusotto, Phys.
Rev. A \textbf{80}, 043603 (2009).

\bibitem{Lahav}   O. Lahav \emph{et al.}, Phys. Rev. Lett.
\textbf{105}, 240401 (2010).

\bibitem{Rousseaux2}    G. Rousseaux \emph{et al.}, New J. Phys.
\textbf{12}, 095018 (2010).

\bibitem{Belgiorno} F. Belgiorno \emph{et al.}, Phys. Rev. Lett.
\textbf{105}, 203901 (2010).

\bibitem{Fouxon}    I. Fouxon \emph{et al.}, Europhys. Lett.
\textbf{92}, 14002 (2010).

\bibitem{Weinfurtner}   S. Weinfurtner \emph{et al.}, Phys. Rev.
Lett. \textbf{106}, 021302 (2011).

\bibitem{Zakharov}  V. E. Zakharov and A. B. Shabat, Sov. Phys.
JETP \textbf{34}, 62 (1972).

\bibitem{Madelung}  E. Madelung, Z. Phys. \textbf{40}, 322 (1927).

\bibitem{Marburger} J. H. Marburger, Prog. Quant. Electron.
\textbf{4}, 35 (1975).

\bibitem{El}    G. A. El \emph{et al.}, Physica D \textbf{87}, 186
(1995).

\bibitem{Fleischer} W. Wan, S. Jia, J. W. Fleischer, Nat. Phys.
\textbf{3}, 46 (2007).

\bibitem{Jia}   S. Jia, W. Wan, J. W. Fleischer, Phys. Rev. Lett.
\textbf{99}, 223901 (2007).

\bibitem{Barsi} C. Barsi \emph{et al.}, Opt. Lett. \textbf{32},
2930 (2007).

\bibitem{Fleurov}   V.Fleurov and A.Soffer, Europhys. Lett.
\textbf{72}, 287 (2005).

\bibitem{Dekel1}    G. Dekel \emph{et al.}, Phys. Rev. A
\textbf{75}, 043617 (2007).

\bibitem{Dekel2}    G. Dekel \emph{et al.}, Phys. Rev. A
\textbf{81}, 063638 (2010).

\bibitem{Barak} A. Barak \emph{et al.}, Phys. Rev. Lett.
\textbf{100}, 153901 (2008).

\bibitem{Wan}   W. J. Wan, S. Muenzel, and J. W. Fleischer, Phys.
Rev. Lett. \textbf{104}, 073903 (2010).

\bibitem{Cohen} E. Cohen \emph{et al.}, arXiv: 1107.0627.

\bibitem{Hakim} V. Hakim, Phys. Rev. E \textbf{55}, 2835 (1997).

\bibitem{Leszczyszyn}   A. M. Leszczyszyn \emph{et al.}, Phys. Rev.
A \textbf{79}, 063608 (2009).

\bibitem{Gladush}   Yu. G. Gladush \emph{et al.}, Phys. Rev. A
\textbf{75}, 033619 (2007).

\bibitem{Fleurov2} V. Fleurov and R. Schilling, Phys. Rev. A{\bf 85},
045602 (2012).


\end{thebibliography}
\end{document}